## II.4 - Non-Hermitian physics in magnetic systems

*Tao Yu[1] and Jinwei Rao[2]*

[1] Huazhong University of Science and Technology, China
[2] ShanghaiTech University, China

**Status**

Recent studies demonstrated that the dissipation is not always detrimental but can be exploited to achieve non-Hermitian topological phases with unexpected functionalities for potential applications. Magnonic devices are low-energy consumption instruments for reprogramable logic, non-reciprocal communication, and non-volatile memory functionalities, in which engineering of dissipation may lead to several non-Hermitian topological phases of magnons, with spin functionalities possibly superior to their electronic, acoustic, optic, and mechanic counterparts [1,2], such as strongly enhanced magnonic frequency combs, magnon or spin-wave amplification, scattering enhancement of magnons, (quantum) sensing with unprecedented sensitivity, magnon accumulation, perfect absorption of microwaves, and magnon bound states in the continuum.

So far, researchers have developed unified approaches based on the master equations and Green-function approach as well for the non-Hermitian topological phases in magnonics. Theoretical and experimental progresses towards engineering dissipation have achieved the non-Hermitian topological phases in magnonic systems, including exceptional points (EPs) [3-5], exceptional nodal phases [6], non-Hermitian Su-Schrieffer-Heeger (SSH) model [7], and non-Hermitian skin effect [8,9] (refer to Fig. 1 for an overview). The EPs in the parameter space refer to the coalescence of eigenvalues and eigenvectors of a non-Hermitian Hamiltonian matrix. Studies have developed two pathways to achieve such EPs with magnons. One is to manipulate either the photon-magnon coupling strength or the gain or loss of a subsystem in cavity magnonics [3,4]. The other one is to design magnetic heterostructures with ferromagnetic and nonmagnetic metal layers [5]. When the complex frequency is a function of the continuous wave vector, the EPs are protected by non-zero topological index, referred to as the exceptional nodal phase. There is a bulk Fermi arc that characterizes the branch cut of the complex energy, which was predicted in magnetic junctions [6]. The generalization of SSH model to the non-Hermitian magnetic system in terms of an array of spin-torque oscillators promises the topological magnonic lasing edge modes, which can be excited by spin current injection [7]. Interesting non-Hermitian skin effects are predicted recently in an array of magnetic wires coupled with the magnetic films via the dipolar interaction, in which the combination of chirality and dissipation of traveling waves drives all the modes to one edge [8]. Such non-Hermitian skin modes are predicted as well in a van der Waals ferromagnetic monolayer honeycomb lattice [9], which is driven by Dzyaloshinskii-Moriya interaction and non-local magnetic dissipation. Strong accumulation of magnon modes at one boundary significantly enhances the sensitivity in detection of small signals [8].

Most predictions in the exceptional nodal phases [6], non-Hermitian SSH model [7], and non-Hermitian skin effect [8,9] with unique functionalities beyond the Hermitian scenario still await the experimental confirmation in the future.

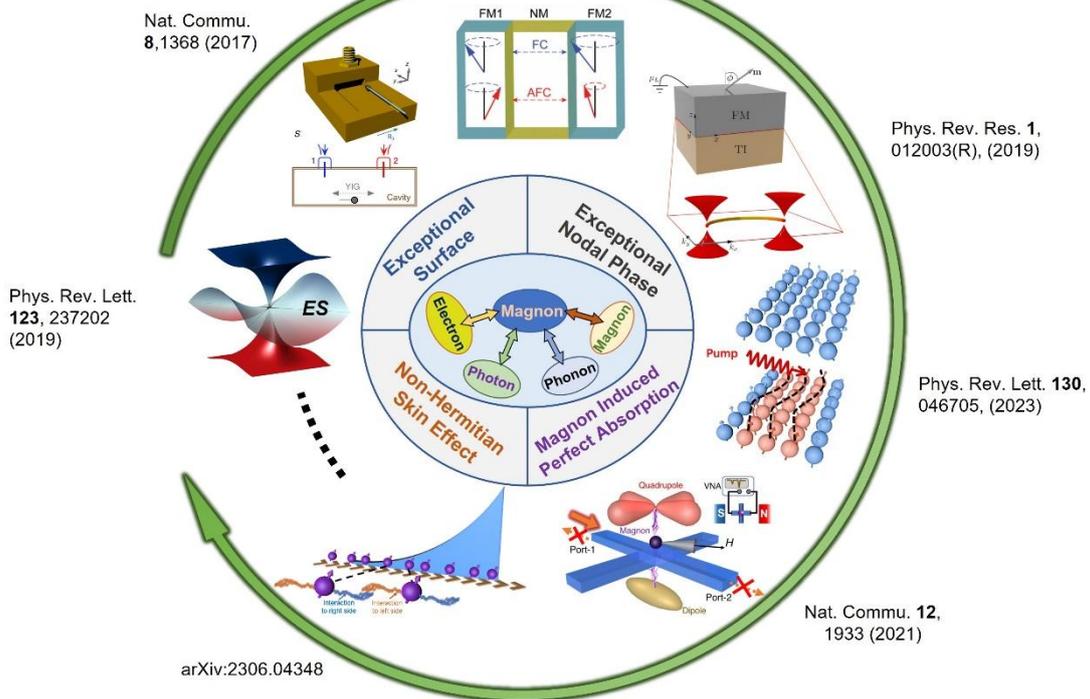

Figure 1. An overview of the present status of non-Hermitian topological magnonics.

**Current and Future Challenges**

**Materials:** To realize and exploit the predictions of novel hybridized magnon modes in non-Hermitian topological magnonics, magnetic materials with high-quality factors and precisely controlled magnetic parameters are favorable. So far, the most commonly used ferromagnetic materials for these purposes are yttrium iron garnet (YIG), CoFeB, and permalloy (FeNi), because of their excellent magnetic properties. The acoustic property of YIG is excellent as well, and its nanostructures such as ultrathin films and nanowires become available only recently. But high-quality YIG appears to be only compatible with thick gadolinium-gallium-garnet (GGG) substrate. Van der Waals magnet is also the material choice for non-Hermitian topological phases [9], but the convincing magnon transport experiment remains wanting. Nevertheless, it appears to be difficult to globally or locally control the interaction and dissipation when stacking the magnetic films to heterostructures such as with other nonmagnetic layers [5], e.g., copper, platinum and magnesium oxide, or with the magnetic nanowire array [8].

**Fabrication of devices:** Loading YIG spheres of millimeter size into the cavity appears on the one hand to be the most convenient way to implement the EPs, but it encounters the difficulty to be fabricated in nanostructures on a chip. The proposal for the other non-Hermitian topological phases is still lacking in this approach. On the other hand, to fabricate nanoscale magnetic heterostructures to realize various non-Hermitian topological phases, one of the challenges is to precisely control the configurations of different layers or nanowires and to guarantee the interfacial quality between them. Both factors demand a very high-quality film growth technique. Moreover, the scalability of magnonic devices to larger sizes and higher frequencies is also a technical challenge that needs to be overcome.

**Gain of magnons:** The realization of gain for magnons is an important design parameter, but raises challenges since it implies "negative Gilbert damping", which is, however, desired for exceptional points and other non-Hermitian topological phases in several cases. On the one hand, spin-polarized electrical currents can induce spin transfer torque to magnetization, leading to amplification of magnonic signals (section I.3). On the other hand, with a feedback loop the output signal can be fed back into the input, which results in the amplification of spin waves. However, several technical challenges follow as well. It is still difficult to completely overcome the damping rate of convenient metallic and insulating magnetic materials; both the excitation/detection and amplification of magnons require suitable electrical circuits; the nonlinear effects pose a challenge in achieving gain in magnonic devices, because, in some cases, the amplification of magnons leads to the generation of higher harmonics or sidebands, which interferes with the desired signals. While research in this direction encounters many obstacles, even small breakthroughs can have significant benefits in device applications.

**Advances in Science and Technology to Meet Challenges**

In order to progress in non-Hermitian topological magnonics, more experimental efforts are sorely needed. While significant advances in material growth, micro-nano fabrication techniques, and magnon gain may be challenging in the short term, there exist alternative methods that can help to overcome current technical challenges as well.

One feasible way is to develop techniques that can independently manipulate the properties of ferromagnetic nanostructures, for example by using bias voltage (section I.1), field gradient, thermal gradient, laser or microwave pump (sections I.3 and I.7). These external drives can excite spin currents or magnon flows, which can transfer energy and torques among different magnetic layers or spatial positions. By controlling the flow of these spin currents and magnons, it may be possible to effectively engineer the dynamics in non-Hermitian magnonic systems.

The opportunity indeed arises in the nonlinear regime (section I.4). A recent effort achieves a pump-induced magnon mode in a magnet when loaded in a microwave waveguide and driven by a strong microwave pump [10]. This magnon mode displays a high level of tunability when driven to the nonlinear regime, and holds the potential to overcome current technological challenges [10]. Via tuning this mode, the giant enhancement of magnonic frequency combs at EPs is subsequently observed.

In addition, hybrid systems based on cavity magnonics may offer another feasible solution. Magnon modes in different magnetic materials can indirectly couple with each other on the long range via the mediation of a microwave cavity [1]. In a cavity magnonic system, interfacial quality that is vital for magnetic heterostructures becomes less important. Gains that may need delicate design in spintronic devices can easily be realized in a cavity magnonic system by embedding an amplifier or gain material in the microwave cavity. However, an awkward reality in this approach is that nearly all current researches on non-Hermitian cavity magnonics are implemented from the cavity side. The magnon mode is merely adopted as a tunable high-Q resonance, because the direct operation techniques on magnon mode, especially precise controlling its dissipation and realization of its gain, are lacking. Given this, a significant opportunity is to develop tuning or readout techniques that can access the photon-magnon coupling process from the magnon side. For instance, researchers have developed a method to control the radiation dissipation of magnon mode by using a loop antenna. However, the tuning range of magnon damping is of

MHz, which is orders of magnitude smaller than that of a cavity photon mode for achieving an observable dissipative photon-magnon coupling (~100 MHz).

**Concluding Remarks**

Non-Hermitian topological magnonics is an emerging field that seeks to realize functionalities beyond those achievable in the Hermitian scenario. In this sub-field, dissipations, which are typically considered detrimental, can be harnessed as important resources for engineering system dynamics. Recent researches have unveiled a range of novel phenomena, including exceptional points, exceptional nodal phases, chirality as generalized spin-orbit interaction, non-Hermitian SSH model with magnons, and non-Hermitian skin effect with potential device applications holding new functionalities. However, due to various experimental obstacles, most research and promising phenomena are still at the theoretical proposal stage. There are thereby high opportunities in the future experiments and device applications after an improvement in material growth and micro-nano fabrication techniques.

**Acknowledgements**

T.Y. and J.W.R. are financially funded by National Natural Science Foundation of China under Grant No. 12204306, the Shanghai Pujiang Program (No. 22PJ1410700), and the startup grant of Huazhong University of Science and Technology (Grants No.3004012185 and 3004012198).